# Self-Adaptive Quantum Kernel Principal Components Analysis for Compact Readout of Chemiresistive Sensor Arrays

Zeheng Wang [1,2,†], Timothy van der Laan [2], and Muhammad Usman [1,3]

[1] Data61, CSIRO, Clayton, VIC 3168, Australia

[2] Manufacturing, CSIRO, West Lindfield, NSW 2070, Australia

[3] School of Physics, The University of Melbourne, Parkville, VIC 3010, Australia

† Corresponding author. Email: zenwang@outlook.com

Keywords: Quantum computing; Quantum PCA; Quantum Machine Learning; Chemiresistive sensor; IoT

## Abstract

The rapid growth of Internet of Things (IoT) devices necessitates efficient data compression techniques to handle the vast amounts of data generated by these devices. Chemiresistive sensor arrays (CSAs), a simple-to-fabricate but crucial component in IoT systems, generate large volumes of data due to their simultaneous multi-sensor operations. Classical principal component analysis (cPCA) methods, a common solution to the data compression challenge, face limitations in preserving critical information during dimensionality reduction. In this study, we present self-adaptive quantum kernel (SAQK) PCA as a superior alternative to enhance information retention. Our findings demonstrate that SAQK PCA outperforms cPCA in various back-end machine-learning tasks, especially in low-dimensional scenarios where access to quantum bits is limited. These results highlight the potential of noisy intermediate-scale quantum (NISQ) computers to revolutionize data processing in real-world IoT applications by improving the efficiency and reliability of CSA data compression and readout, despite the current constraints on qubit availability.

## Introduction

Quantum computing (QC) is increasingly recognized as a pivotal solution for computationally intensive problems such as integer factorization[1,2] and quantum system simulations[3,4]. Likewise, it is anticipated that the integration of quantum computing in machine learning (ML) and data processing tasks will offer computational advantages such as speed-up[5], enhanced accuracy[6], and superior robustness[7], made possible by unique quantum properties like superposition and entanglement. These properties allow data to be stored and processed in a potentially high-dimensional quantum space, leading to computation optimization[8]. Quantum machine learning (QML) has already demonstrated high potential in various applications. For instance, a hybrid classical-quantum principal component analysis (PCA) has been applied in drug design[9], while other studies have explored quantum computing applications in finance and enhanced classical machine learning[10,11]. Additionally, it has been theoretically predicted that quantum kernel methods could consistently outperform classical counterparts in modeling data with group structure[8].

However, experimental research leveraging QML to solve real-world problems by identifying group structures in datasets has not yet been extensively carried out, leaving this theoretical work[8] largely unvalidated. Simultaneously, in noisy intermediate-scale quantum (NISQ) systems, practical applications of QML are becoming increasingly feasible[12–15]. Despite recent advancements in high-density QC architectures, the number of logical qubits in QC processors still falls short of demonstrating a significant quantum advantage[16–19]. This gap presents a unique opportunity to apply QML for practical problem-solving, particularly in enhancing data compression and readout compactness, when the number of qubits is a critical limitation. To address these challenges, this study employs a quantum kernel-based algorithm, quantum PCA (qPCA), to compress data and enhance backend data processing and therefore bridges the gap by demonstrating the practical utility of qPCA in IoT data compression, a real-world application.

The experimental data used in this work was obtained from chemiresistive sensor arrays (CSAs), an example of a widely used IoT device due to their fabrication simplicity, high sensitivity, and specificity in detecting chemical changes. However, CSAs generate large volumes of data due to their requirement for multiple sensors to operate simultaneously, posing significant challenges for data compression and processing. While classical compression techniques have been developed[20–23], these methods often struggle with maintaining data integrity during the reduction process, leading to loss of critical information[24]. Our findings demonstrate that qPCA outperforms classical PCA (cPCA) in preserving critical information during dimensionality reduction, leading to more efficient and reliable data modeling.

This manuscript discusses the integration of QML with real-world data challenges, making a case for the adoption of quantum methods in broader contexts beyond theoretical applications. As shown in Fig. 1, in this study, we employ quantum PCA to achieve compact readout of CSA data. We utilized a self-adaptive quantum kernel (SAQK) technique upon the fidelity-based quantum kernel to map readings from seven sensors into quantum state space and then compress the data by reducing the dimension of that space. Following dimensional reduction, we applied various classical ML (CML) algorithms to determine the readout results. Compared to the classical kernel (radio-basis function kernel, RBFK) and two trivial quantum kernels, our SAQK demonstrated lower information loss, leading to enhanced ML-based readout accuracy, which has only been theorized previously[8]. This study is conducted on classical hardware by simulating quantum circuits to evaluate the practical feasibility and advantages of qPCA within the constraints of current NISQ devices. Specifically, our implementation considers a small-scale quantum system of up to 7 qubits, aligning with typical hardware capacities of present-day quantum computers. These simulations provide insights into the potential advantages of quantum-enhanced dimensionality reduction, paving the way for future deployment on real quantum hardware as devices mature.

## Self-adaptive quantum kernel

We propose a SAQK framework to address the real-world challenge of classifying group-structured, classically inseparable sensor data by leveraging adaptive quantum kernel mapping. As shown in Fig. 1, after collecting the data from the classical world, these data's classical features, inherently exhibiting complex group symmetries, are encoded into quantum states through a variationally optimized fiducial state. During this procedure, a self-adaptive layer (a variational quantum circuit consisting of parameterized $R_x$ and $R_z$ gates, see Method) dynamically adjusts the quantum state representation to align with the data's group structure, ensuring a high-fidelity mapping into a covariant quantum space. During the covariant mapping, unitary operations $U_F(\boldsymbol{x})$ embed the classical data into a quantum Hilbert space, preserving group symmetries while capturing non-linear relationships between data points. This intermediate quantum space facilitates inseparable data but retains invariance under group transformations.

Subsequently, a quantum kernel $U_K(\boldsymbol{x}, \widetilde{\boldsymbol{x}})$ is constructed by computing pairwise fidelities between quantum states, effectively forming a high-dimensional kernel-defined space where class-wise patterns become increasingly separable. In this final kernel-defined space, the originally inseparable data points are separable, enabling accurate classification using classical machine learning models. This framework demonstrates how the SAQK effectively leverages simple universal rotation gates and covariant quantum mapping to transform real-world, group-structured, and inseparable data into a separable form suitable for robust classification.

## Quantum and classical kernel matrices

Two different, trivial quantum kernels and one classical kernel (RBF-based) are benchmarked with the proposed SAQK in this study (different quantum kernels use different data mapping strategies to map the classical data to the quantum space, see Supplementary Figs. 1, 2, and 3 for details). Fig. 2 illustrates the distinctions between quantum and classical kernel matrices, along with their associated data embeddings in low-dimensional (2D) feature spaces, highlighting their varying capabilities in capturing data structures. Fig. 2(a), the Pauli-X kernel, exhibits minimal redundancy but also minimal correlation, as reflected in its uniform appearance, rendering it ineffective at capturing nuanced data relationships. In contrast, the ZZ-Map kernel (Fig. 2(b)) introduces moderate correlation, with darker blocks indicating stronger relationships within subsets of data. The SAQK (Fig. 2(c)), however, demonstrates an advanced quantum approach, achieving intricate patterns of both correlation and redundancy, which align well with complex data structures. This contrasts with the classical RBF kernel (Fig. 2(d)), which captures broader, simpler relationships with more redundancy, evident in its uniform dark stripes.

The 2D data embeddings further emphasize these differences. The ZZ-Map (Fig. 2(f)) and SAQK (Fig. 2(g)) produce separable clusters that closely align with intrinsic data structures, while the RBF kernel (Fig. 2(h)) generates denser and more uniform groupings, potentially overlooking finer details and preserving redundancy unnecessarily. These results highlight the SAQK kernel's ability to capture subtle, non-linear relationships in high-dimensional feature spaces, making it particularly effective for preserving intricate patterns. In contrast, classical kernels like RBF tend to emphasize dominant global trends, limiting their ability to represent data with complex group structures[25]. This comparison underscores the unique advantages of quantum kernels, especially SAQK, in scenarios involving intricate or group-structured data distributions. Fig. 2(i) illustrates an example of the quantum state vector produced by the SAQK feature map, with all qubit states aligning near maximally coherent configurations. This coherence highlights the effective exploitation of quantum advantage in feature mapping, facilitating an enhanced representation of complex data relationships[26].

It should be noted that, however, the ultimate choice of kernel for a given task should not rely solely on visual inspection; it must be informed by empirical performance metrics, such as classification accuracy, regression error, or other relevant benchmarks in machine learning models.

## Benchmarking SAQK PCA and cPCA

To explore the impact of SAQK qPCA and cPCA on information retention during dimensionality reduction, we evaluated multiple machine learning algorithms for predicting the chemical types detected by the CRS across different samples. The selected models span a diverse range of representative approaches, including linear classifiers, kernel-based methods, decision tree ensembles, neural networks, and high-performance ensemble-learning (EL) methods. By using these algorithms as probes, we systematically assess the efficacy of the kernels in preserving and leveraging critical information. As delineated in Fig. 3(a), data refined through qPCA were more amenable to modeling by these algorithms, evidenced by their generally higher evaluation scores. Nonetheless, it is imperative to acknowledge the conditions under which cPCA-treated data surpassed qPCA's performance. A notable instance is observed with dimensions above 5D for the EL algorithms, where cPCA generally maintained higher scores. This suggests that cPCA can preserve certain intricate non-linear patterns within the original dataset, which the EL algorithms can exploit to garner additional insights. This preservation of complex structures, however, seems to diminish rapidly with further dimensional reduction, as demonstrated by the comparative analysis within the EL subpanels in Fig. 3.

Examining the lower-dimensional performance of linear-based algorithms, such as LR and L-SVM, we also see cPCA perform better than qPCA, as shown in lower panel, Fig. 3(a). Despite cPCA's use of the inherently non-linear RBF kernel, this phenomenon indicates preferential retention of linear-like patterns through the dimensionality reduction process. The qPCA may be overlooking linear patterns during dimensionality reduction hence the lower scores compared to the cPCA. However, we observe that the scores for both cPCA and qPCA methods, as well as LR and L-SVM, are generally low, indicating that the linear-like patterns represent only a small portion of the overall key information. In addition, the low scores may indicate that while the cPCA method preserves linear patterns during the dimensionality reduction, it could be omitting nonlinear patterns in the data. Conversely, the SAQK PCA method is more adept at retaining nonlinear patterns. This suggests that qPCA could be particularly effective for data dimension compression when the critical information within the data predominantly resides in nonlinear patterns.

It is however crucial to again recognize that those linear patterns form only a fragment of the overall information necessary for robust modeling the CSA data. This is substantiated when juxtaposing the performance metrics of linear and non-linear models, the latter typically presents higher scores. While cPCA occasionally demonstrates advantageous outcomes, particularly in partially preserving linear components and certain non-linear patterns beneficial to algorithms like RF in uncompressed space, the overarching readout results narrative attests to the superior comprehensive capability of SAQK PCA.

The quantum advantage becomes more evident when focusing on the high-performance EL algorithms (Fig. 3(b)-(e)) and their corresponding statistics (Fig. 3(f)-(k)). In particular, RF and ET exhibit consistently higher scores during dimensional reduction. The performance of other EL models varies, potentially due to the inherent preferences of specific ML algorithms or the limitations highlighted by the no-free-lunch theorem[27]. Statistical analysis of the best-performing models shows that SAQK PCA achieves significantly lower average collapse rates (decline in scores with dimensional reduction), highlighting its robustness in preserving information.

## Verification by artificial low-entropy data

To further validate the superiority of SAQK PCA over cPCA in maintaining key information of the data, we synthesized a set of artificial low-entropy datasets designed to mimic real-world yet challenging data scenarios. As shown in Fig. 4, the datasets included both linear and nonlinear structures, generated by sampling from Gaussian distributions with strong noise levels and augmented with standardization for consistency. Each dataset consisted of seven features and labels across 300 data points, enabling a direct comparison of information retention during dimensionality reduction from 7D to 2D.

We applied SAQK PCA and cPCA to both linear and nonlinear datasets, followed by machine learning-based evaluations using the high-performance EL algorithms. The same performance metrics, accuracy, F1 score, and Cohen's Kappa (CK) score were computed to assess classification quality across different reduced dimensions. As shown in Fig. 4(d)-(f), during dimensional reduction, the evaluation scores almost fluctuate at the same level due to the low entropy's nature, but it is still clear that SAQK PCA outperforms cPCA, particularly in nonlinear scenarios, where the SAQK's adaptability ensures better classification outcomes. Statistical analyses, shown in Fig. 4(g)-(l), further highlight the advantage of SAQK PCA in the non-linear dataset by leveraging complex data structures, underscoring its advantage in preserving critical information during dimensionality reduction.

## Discussion

Recent advancements in quantum machine learning theory highlight the efficacy of quantum kernels in classifying data with inherent group structures[8]. Using our experimental dataset, we demonstrate the effectiveness of the SAQK methods and their potential for real-world applications. We provide evidence that SAQK kernel-based PCA offers promising solutions for complex data compression problems with non-obvious relationships governed by group structures. Particularly for processing from high to low dimensions, the proposed qPCA method shows promise, as applying quantum data processing using qubits has become feasible in the NISQ era. Looking ahead, the scalability of SAQK PCA to larger datasets and even higher-dimensional feature spaces depends on advancements in quantum hardware: As quantum devices evolve, the ability to process datasets with significantly more features will increase, enabling broader application of SAQK PCA to complex IoT systems.

While this study highlights SAQK PCA's potential advantages over cPCA in our framework, it is yet to be determined whether qPCA universally outperforms all kernel-based cPCA methods. It should be noted that a carefully crafted kernel or tailored post-processing algorithm could enhance cPCA's ability to preserve critical information during dimensionality reduction. Moreover, implementing cPCA is straightforward with established algorithms and software support, allowing easy application with conventional computing resources. In contrast, contemporary qPCA requires higher computational demands and specialized quantum computing resources, reflecting its developmental stage. Additionally, other dimensionality reduction techniques may prove more suitable for specific tasks where the limitations of qPCA render it less effective (see Supplementary Table 1).

This study is based solely on simulations under the assumption that the qubits used for processing data are ideal. While the demonstrated ability of qPCA to retain more informative features highlights its potential as a promising technique for advanced data compression in IoT systems, the presence of noisy qubits could significantly degrade its performance[8]. Nevertheless, the robustness of SAQK PCA to noisy data suggests its potential to adapt effectively in practical scenarios involving NISQ devices. Future research should therefore focus on evaluating the application of qPCA across diverse data types and tasks, as well as optimizing qPCA kernels to accommodate varying data structures in quantum computers of the NISQ era by incorporating quantum noise models to further assess the real-world performance. Additionally, exploring the deployment and performance of quantum machine learning models, including qPCA, on commercially available quantum hardware presents an exciting avenue for further investigation.

## Summary

In conclusion, this study demonstrates the advantage of SAQK PCA over cPCA for dimensionality reduction in IoT applications, particularly for datasets with complex, non-linear relationships governed by group structures. By leveraging self-adaptive quantum kernels, SAQK PCA effectively retains critical information during compression, achieving consistently higher readout performance across diverse machine learning models. These findings experimentally validate theoretical predictions of quantum-enhanced dimensionality reduction, even on NISQ devices with limited qubits. While challenges such as computational overhead remain, this work highlights the potential of quantum approaches to advance data compression techniques and opens avenues for further optimization and broader applications.

## Methods

**Experimental data**: The chemiresistive sensor array used in this study consisted of 17 sensors, each based on gold nanoparticle films deposited on interdigitated electrodes. The gold nanoparticles were functionalized with different thiols to impart partial selectivity to specific analytes[28]. For instance, the 1,10-decanedithiol functionalized sensor (1-10-DDT) exhibited a relatively weak response to naphthalene compared to the 1-heptanethiol functionalized sensor (1-HEPTT). This functionalization allowed the array to achieve differential responses to various chemical components.

The sensor array was exposed to a comprehensive set of chemical mixtures consisting of benzene (B), toluene (T), ethylbenzene (E), p-xylene (X), naphthalene (N), and a mixture of interferants (I) comprising organics that could potentially interfere with the sensor responses to the BTEX analytes. A full factorial Design of Experiment was employed, ensuring all possible combinations of these components were prepared and tested. Each mixture was exposed to the sensor array 12 times, and the maximum relative resistance change ($\Delta R/R_0$) for each sensor was recorded during each exposure[29].

We sourced our data from the dataset of the resistance changes. The sensor array naturally introduces some level of noise into the measurements. To retain the authenticity of the real-world dataset, no explicit noise filtering was applied. This dataset comprises readings from 17 sensors across 852 experiments (data shape: 852×17), with each 17-dimensional data point linked to one of 66 labels identifying the detected chemical. Due to the qubit limitations inherent in NISQ systems, we initially narrowed our focus to data from seven specific sensors (4-BBM, 1-2-BDMT, MOB, 3-ETP, 4-MBT, 4-CBT), resulting in a refined dataset of 852 experiments, each with 7-dimensional readings (selected data shape: 852×7), by the technique we proposed previously[28]. Each data item consisted of these 7D readings paired with a label. The whole procedure of the data processing and analysis can be found in Fig. 5. The experimental setup for collecting the CSA data is briefly illustrated in the left panel of Fig. 1 and more details about the sensor's fabrication and the wet experiments can be found in our previous work[29].

For data compression, we employed kernel PCA utilizing four distinct kernel types: the Pauli-Z kernel, the ZZ-Feature kernel, the classical RBF kernel, and the RBFK.

The quantum kernel-based PCA implementation in this study uses a maximum of 7 qubits, corresponding to the dimensionality of the dataset (D=7). These simulations, realized by Qiskit packages, mimic realistic quantum circuit behaviors, including unitary operations and kernel evaluations, as would be implemented on current quantum devices. While this work uses classical simulations, the findings are directly applicable to NISQ-era quantum computers.

**Classical PCA**: For RBFK PCA, the kernel is defined by the Gaussian function:

$$K(x_i, x_j) = \exp\left(-\frac{|x_i - x_j|^2}{2\sigma^2}\right) \quad (1)$$

$\|x_i - x_j\|$ is the Euclidean distance between data points $x_i$ and $x_j$, and $\sigma$ is a free parameter that controls the width of the Gaussian.

Applying these kernel methods, we performed dimensionality reduction on the 7D data, mapping it into a feature space via a kernel function $K$. In this space, we computed the covariance matrix $C$ defined by the kernel as:

$$C = \frac{1}{n}\sum_{i=1}^{n}(K \cdot \phi(x_i))(K \cdot \phi(x_i))^T \quad (2)$$

where $n$ is the number of data samples, $\phi(x_i)$ is the implicit mapping of the data point $x_i$ by the kernel function $K$, and $T$ denotes the transpose operation. We then solved the eigenvalue problem:

$$C \cdot v = \lambda \cdot v \quad (3)$$

where $v$ are the eigenvectors and $\lambda$ are the eigenvalues. The eigenvectors corresponding to the largest eigenvalues give us the principal components in the feature space. By selecting the top $k$ eigenvectors, we can form a reduced feature space of dimension $k$, where $k$ is less than the original dimensionality of the data. For our purposes, we chose $k = 6, 5, 4,$ and $3$ to obtain a 6D, 5D, 4D, and 3D representation of the data, respectively. In this reduced feature space, the transformed data points $\hat{x}_i$ are given by:

$$\hat{x}_i = [v_1^T \cdot \phi(x_i), v_2^T \cdot \phi(x_i), \ldots, v_k^T \cdot \phi(x_i)] \quad (4)$$

This transformation is expected to retain the essential features of the original data and to be suitable for processing within a NISQ environment due to the lower dimensionality. It is now evident that different mapping strategies, i.e. different $K$ and $\phi(x_i)$, will lead to different transformed data points $\hat{x}_i$.

**Quantum PCA**: Kernel PCA transforms the data into a higher dimensional space defined by a kernel function $K$, where $K(x_i, x_j)$ represents the similarity between two points in the original space. The transformation is given by:

$$\Phi(x) = K(\cdot, x) \quad (5)$$

In the case of QFK PCA, we use the quantum fidelity between quantum states, which is defined as:

$$F(\rho_i(x_i), \rho_j(x_j)) = \left(\mathrm{Tr}\sqrt{\sqrt{\rho_i(x_i)}\rho_j(x_j)\sqrt{\rho_i(x_i)}}\right)^2 \quad (6)$$

Here, $\rho_i(x_i)$ and $\rho_j(x_j)$ are density matrices representing the quantum states mapped from the data. In this work, we adopted Pauli feature map function to realize the mapping from classical space to quantum state space[30]. The qubits were initialized to $|0\rangle^7$ states to encode the CSA data. The fidelity measures the closeness between two quantum states, which in our context, serves as a measure of similarity between data points when transformed to quantum state space. We can define our kernel matrix $K$ where each element $K(x_i, x_j)$ is the fidelity between states $\rho_i$ and $\rho_j$:

$$K(x_i, x_j) = F(\rho_i, \rho_j) = \left(\mathrm{Tr}\sqrt{\sqrt{\rho_i}\rho_j\sqrt{\rho_i}}\right)^2 \quad (7)$$

The qPCA's procedure is illustrated in Fig. 5, where the quantum kernel estimation (QKE) was realized by the sampling function of Qiskit[31]. Following the QKE, the classical mapping algorithm was used to calculate the principal components and to realize the feature mapping, as shown in Fig. 5.

**SAQK method:** In this study, we develop a self-adaptive quantum kernel method to enhance the compression efficiency of group-structured, classically inseparable data. Our approach is inspired by the theory presented in Ref[8]. Notably, two key distinctions set our approach apart: First, to incorporate the data's behaviors, we choose Pauli-Z feature map, $U_F(x) = \prod_{i=0}^{n} R_z(x_i)$, which encodes classical data into quantum states using Z-axis rotations proportional to the input features in a more straightforward way; Second, according to our experience, we introduce the trainable parameters in our self-adaptive quantum kernel framework via $U_V^\theta(\mathbf{x}, \theta) = \prod_{i=0}^n R_x(x_i, \theta_i)$ rotations.

By combining the $R_x$ and the $R_z$ gates, the approximately universal operations can also be approached when realizing the covariant mapping without adding more complexity in quantum mapping as in Ref[8]. The self-adaptive training process and its corresponding quantum circuit are detailed in Supplementary Algorithm 1 and Supplementary Fig. 1. The detailed theory of the trainable kernel can be found in Ref[8] and Ref[31].

**t-SNE embedding for data visualization:** t-SNE was applied using the Scikit-Learn implementation with default parameters (perplexity $p = 30$, learning rate $lr = 200$, and iterations $t = 1000$) to visualize the data in 2D. These settings provide a consistent baseline for comparing the structures captured by qPCA and cPCA, avoiding potential biases introduced by hyperparameter tuning.

**Machine learning models**: Following dimensionality reduction, the information retention in the compressed data was evaluated using a diverse set of ML models, categorized into three groups: linear models, non-linear models, and EL models. The evaluation included:

1. Linear Models: Logistic regressor (LR) and linear kernel support vector machine (L-SVM).
2. Non-Linear Models: Radial basis function support vector machine (RBF-SVM), k-nearest neighbors' classifier (KNN), Gaussian naive bayes classifier (NB), and neural networks based on multi-layer perceptron classifier (MLP).
3. Ensemble Learning Models: Bagging-based Random forests (RF) and extremely randomized trees (ET), boosting-based gradient boosting classifiers (GBC) and extreme gradient boosting (XGB).

These models encompass a comprehensive range of methodologies, from linear classifiers and kernel-based techniques to decision tree ensembles and neural networks. This diversity ensures a robust assessment of how well each algorithm leverages the retained information in reduced data spaces (7D to 2D). By employing both cPCA and qPCA, the study provides an extensive performance analysis of these ML models across varying dimensions and data complexities.

To consolidate the strengths of all CML models, we also employed EL frameworks. These frameworks were utilized to integrate the results from the individual base models – The final prediction was made by all the individual models by bagging or boosting mechanism[32].

For the evaluation, we extracted key metrics — Accuracy, F1 score, and Cohen's Kappa (CK) score[33] — along with their rate of change across different dimensions, to illustrate the outcomes from the two ensemble learning frameworks. The datasets compressed via RBFK and SAQK PCA were randomly split into training (80%) and testing (20%) sets, with stratified sampling. To ensure robustness in our results, we repeated this data splitting process 10 times, calculating the average evaluation scores over these iterations for each algorithm.

For the implementation, we used Python 3.10, with ML models sourced from the Scikit-Learn package (version 1.4.0) and quantum ML models from Qiskit 0.44.1 and Qiskit-Machine-Learning 0.6.0[31]. All ML models' settings used the default settings directly. The kernels employed for this study were the fidelity kernel for quantum analysis and the RBF kernel for classical analysis. The code was executed on a computer equipped with a Ryzen 5600G CPU and 64Gb of memory.

**Low-entropy data synthesis**: Low-entropy data synthesis was designed to generate datasets with controlled complexity and reduced information entropy for benchmarking machine learning models in a way similar to the worst real-world case. This process involves generating both linearly and non-linearly data through parametric equations and labeling the data by step functions. The synthesized data maintains class balance and allows controlled injection of noise and non-linearity.

Linear data is generated using a weighted combination of features $\mathbf{X} \cdot \mathbf{w}$ and Gaussian noise $\epsilon$. For $n$ samples and $d$ features:

$$\mathbf{X} \sim \mathbb{N}(0,1)^{n \times d}, \quad \mathbf{w} \sim \mathbb{N}(0,1)^d \quad (8)$$

Note the weights $\mathbf{w}$ are also sampled from the Gaussian distribution $\mathbb{N}$ to reduce the entropy of the data and emulate the real-world cases. The linear combination is calculated as:

$$\mathbf{y} = \mathbf{w} \cdot \mathbf{X} + \epsilon, \quad \epsilon \sim \mathcal{N}(0, \sigma^2) \quad (9)$$

Classification labels are assigned based on a threshold 0:

$$y_i = f(x_i) = \begin{cases} 0, & f(x_i) < 0 \\ 1, & otherwise \end{cases} \quad (10)$$

Non-linear data introduces complex feature interactions using trigonometric and polynomial transformations:

$$\mathbf{y} = sin(\mathbf{w} \cdot \mathbf{X}) + cos(\mathbf{w} \cdot \mathbf{X}) + e^{-\mathbf{w} \cdot \mathbf{X}} + ln(|\mathbf{X}| + 1) + \epsilon, \quad \epsilon \sim \mathcal{N}(0, \sigma^2) \#(11)$$

Labels are assigned using the same step function as the linear case. For both cases, the noise $\epsilon$ is sampled from a Gaussian distribution with 0 mean and a variance of 0.1.

This noise is added directly to the feature combinations $\mathbf{y}$ in both linear and non-linear data synthesis processes. In the linear case, it perturbs the weighted sum of the features, shifting the decision boundary. In the non-linear case, it introduces random variations to the complex feature interactions, further mimicking real-world uncertainties. To ensure consistent scaling and compatibility across machine learning models, all features were standardized using Scikit-Learn's StandardScaler method, which normalizes the data by removing the mean and scaling to unit variance.

## Acknowledgements

This work was partially supported by CSIRO Impossible-Without-You program.

## References

1. Jiang, S., Britt, K. A., McCaskey, A. J., Humble, T. S. & Kais, S. Quantum Annealing for Prime Factorization. *Sci. Rep.* **8**, 17667 (2018).
2. Borders, W. A. *et al.* Integer factorization using stochastic magnetic tunnel junctions. *Nature* **573**, 390–393 (2019).
3. Boghosian, B. M. & Taylor, W. Simulating quantum mechanics on a quantum computer. *Phys. Nonlinear Phenom.* **120**, 30–42 (1998).
4. Zalka, C. Simulating quantum systems on a quantum computer. *Proc. R. Soc. Lond. Ser. Math. Phys. Eng. Sci.* **454**, 313–322 (1998).
5. Biamonte, J. *et al.* Quantum machine learning. *Nature* **549**, 195–202 (2017).
6. Cerezo, M., Verdon, G., Huang, H.-Y., Cincio, L. & Coles, P. J. Challenges and opportunities in quantum machine learning. *Nat. Comput. Sci.* **2**, 567–576 (2022).
7. West, M. T. *et al.* Towards quantum enhanced adversarial robustness in machine learning. *Nat. Mach. Intell.* **5**, 581–589 (2023).
8. Glick, J. R. *et al.* Covariant quantum kernels for data with group structure. *Nat. Phys.* (2024) doi:10.1038/s41567-023-02340-9.
9. Batra, K. *et al.* Quantum Machine Learning Algorithms for Drug Discovery Applications. *J. Chem. Inf. Model.* **61**, 2641–2647 (2021).
10. Orús, R., Mugel, S. & Lizaso, E. Quantum computing for finance: Overview and prospects.

*Rev. Phys.* **4**, 100028 (2019).
11. Perdomo-Ortiz, A., Benedetti, M., Realpe-Gómez, J. & Biswas, R. Opportunities and challenges for quantum-assisted machine learning in near-term quantum computers. *Quantum Sci. Technol.* **3**, 030502 (2018).
12. Cheng, B. *et al.* Noisy intermediate-scale quantum computers. *Front. Phys.* **18**, 21308 (2023).
13. Bharti, K. *et al.* Noisy intermediate-scale quantum algorithms. *Rev. Mod. Phys.* **94**, 015004 (2022).
14. West, M. T., Sevior, M. & Usman, M. Boosted Ensembles of Qubit and Continuous Variable Quantum Support Vector Machines for B Meson Flavor Tagging. *Adv. Quantum Technol.* **6**, 2300130 (2023).
15. West, M. T. *et al.* Drastic Circuit Depth Reductions with Preserved Adversarial Robustness by Approximate Encoding for Quantum Machine Learning. *Intell. Comput.* **3**, 0100 (2024).
16. Strikis, A., Benjamin, S. C. & Brown, B. J. Quantum Computing is Scalable on a Planar Array of Qubits with Fabrication Defects. *Phys. Rev. Appl.* **19**, 064081 (2023).
17. Liu, Y., Guan, S., Luo, J. & Li, S. Progress of Gate-Defined Semiconductor Spin Qubit: Host Materials and Device Geometries. *Adv. Funct. Mater.* 2304725 (2024) doi:10.1002/adfm.202304725.
18. Dalzell, A. M., Harrow, A. W., Koh, D. E. & La Placa, R. L. How many qubits are needed for quantum computational supremacy? *Quantum* **4**, 264 (2020).
19. Harrow, A. W. & Montanaro, A. Quantum computational supremacy. *Nature* **549**, 203–209 (2017).
20. Li, S., Xu, L. D. & Wang, X. Compressed Sensing Signal and Data Acquisition in Wireless Sensor Networks and Internet of Things. *IEEE Trans. Ind. Inform.* **9**, 2177–2186 (2013).
21. Pudlewski, S., Prasanna, A. & Melodia, T. Compressed-Sensing-Enabled Video Streaming for Wireless Multimedia Sensor Networks. *IEEE Trans. Mob. Comput.* **11**, 1060–1072 (2012).
22. Fazel, F., Fazel, M. & Stojanovic, M. Random Access Compressed Sensing for Energy-Efficient Underwater Sensor Networks. *IEEE J. Sel. Areas Commun.* **29**, 1660–1670 (2011).
23. Palopoli, L., Passerone, R. & Rizano, T. Scalable Offline Optimization of Industrial Wireless Sensor Networks. *IEEE Trans. Ind. Inform.* **7**, 328–339 (2011).
24. Nassra, I. & Capella, J. V. Data compression techniques in IoT-enabled wireless body sensor networks: A systematic literature review and research trends for QoS improvement. *Internet Things* **23**, 100806 (2023).
25. Ghukasyan, A., Baker, J. S., Goktas, O., Carrasquilla, J. & Radha, S. K. Quantum-Classical Multiple Kernel Learning. Preprint at https://doi.org/10.48550/ARXIV.2305.17707 (2023).
26. Streltsov, A., Adesso, G. & Plenio, M. B. *Colloquium*: Quantum coherence as a resource. *Rev. Mod. Phys.* **89**, 041003 (2017).
27. Wolpert, D. H. & Macready, W. G. No free lunch theorems for optimization. *IEEE Trans. Evol. Comput.* **1**, 67–82 (1997).
28. Wang, Z., Cooper, J. S., Usman, M. & Van Der Laan, T. Blue and Green-Mode Energy-Efficient Nanoparticle-Based Chemiresistive Sensor Array Realized by Rapid Ensemble Learning. *ACS Appl. Nano Mater.* **7**, 24437–24446 (2024).
29. Cooper, J. S. *et al.* Quantifying BTEX in aqueous solutions with potentially interfering hydrocarbons using a partially selective sensor array. *The Analyst* **140**, 3233–3238 (2015).
30. Havlíček, V. *et al.* Supervised learning with quantum-enhanced feature spaces. *Nature* **567**, 209–212 (2019).
31. Treinish, M. Qiskit/qiskit-metapackage: Qiskit 0.44.0. Zenodo https://doi.org/10.5281/ZENODO.2573505 (2023).
32. Mienye, I. D. & Sun, Y. A Survey of Ensemble Learning: Concepts, Algorithms, Applications, and Prospects. *IEEE Access* **10**, 99129–99149 (2022).
33. Vieira, S. M., Kaymak, U. & Sousa, J. M. C. Cohen's kappa coefficient as a performance measure for feature selection. in *International Conference on Fuzzy Systems* 1–8 (IEEE, Barcelona, Spain, 2010). doi:10.1109/FUZZY.2010.5584447.

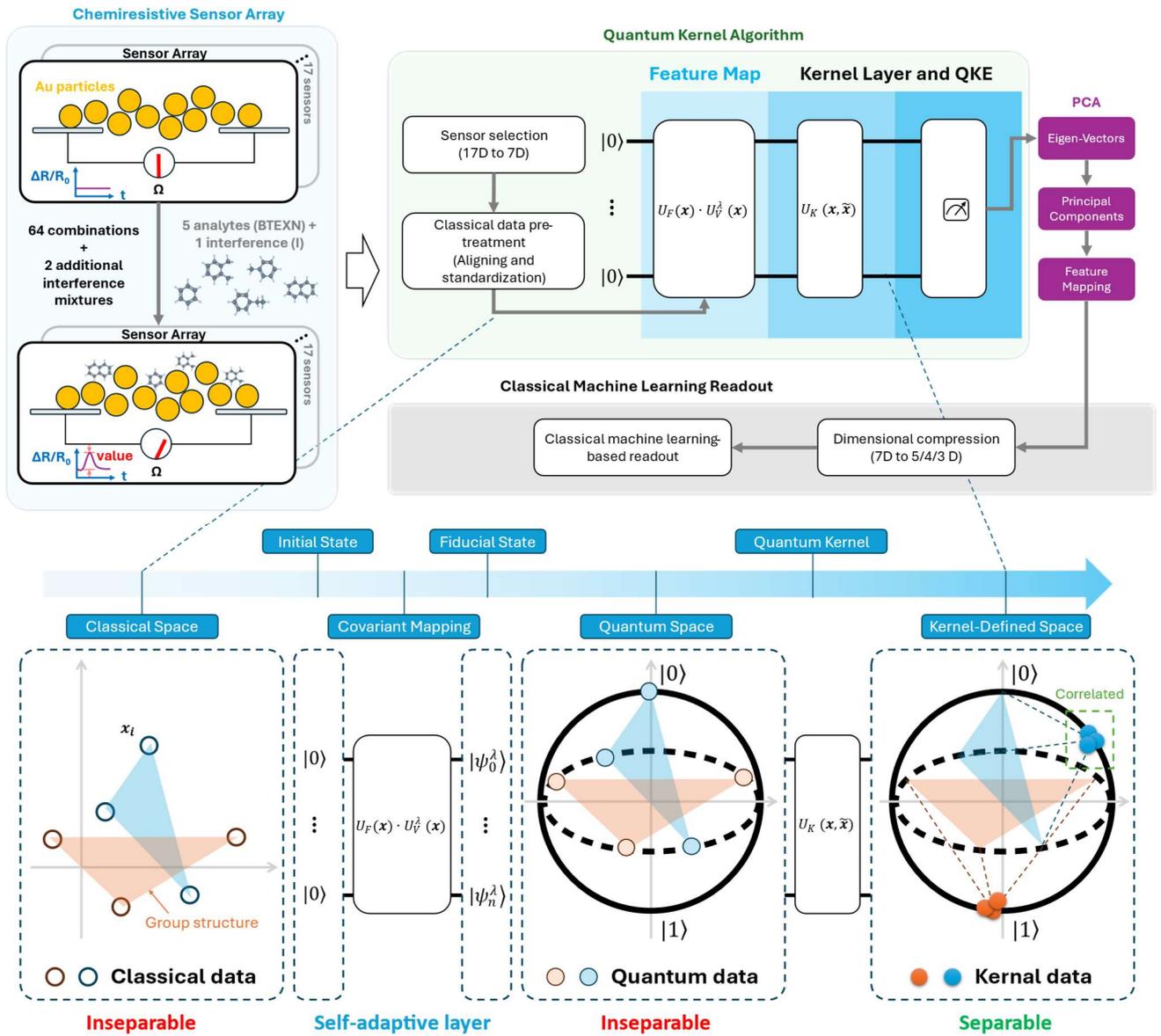

Fig. 1: Schematic illustration of the data processing procedure, emphasizing self-adaptive quantum kernel (SAQK) PCA. The chemiresistive sensor array (CSA) consists of 17 sensors, and data were collected from the 7 most sensitive sensors for detecting 5 analytes and 1 interference mixture (see Methods). The quantum kernel algorithm comprises the covariant feature mapping $U_F(x) \cdot U_V^\lambda(x)$ which adaptively transforms inseparable classical data into a quantum space, and kernel estimation $U_K(x, \tilde{x})$ using Qiskit's sampling function to compute quantum fidelities for kernel matrices. Note that the $U_V^\lambda(x)$ comprises a static feature map (Pauli-Z) and a variational layer to realize the covariant mapping and the kernel alignment (see Supplementary Algorithm 1). Quantum and Classical Principal component analysis (qPCA/cPCA) was then performed to extract principal components and achieve dimensionality reduction from 7D to lower dimensions (6D to 2D). The data with reduced dimensions is then used for classical machine learning (ML)-based readout for benchmarking the qPCA and cPCA efficiency.

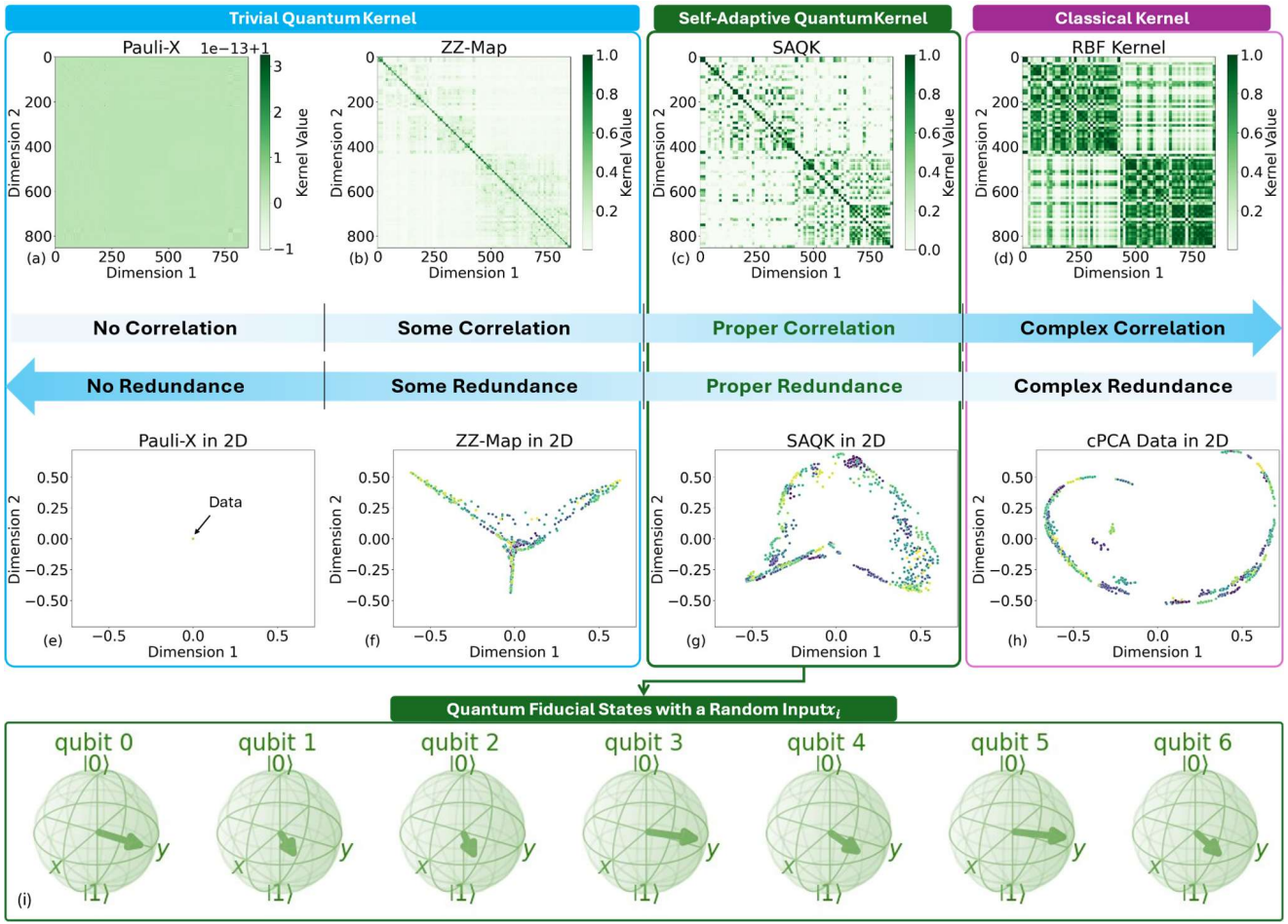

Fig. 2: Comparison of quantum and classical kernel matrices and their t-SNE-based 2D embeddings (the algorithm can be found in Supporting Information). This figure benchmarks the proposed SAQK against typical quantum and classical kernels. From (a) to (c), the quantum kernel matrices exhibit increasingly scattered and irregular patterns, indicating the ability to capture more complex correlations in the quantum feature space. Darker regions signify higher similarity between data points, reflecting the redundancy within the kernel space. (d) The classical kernel matrix, constructed using an RBF kernel, shows a more uniform and regular structure with distinct dark stripes, highlighting dominant correlations but also higher redundancy. The differences, particularly in off-diagonal regions, underscore SAQK's ability to capture nuanced, non-linear, group-structured relationships while mitigating redundancy. (e)-(g) Visualizations of the compressed data using t-SNE embeddings after dimensionality reduction with qPCA. The SAQK projection reveals dispersed clusters, reflecting the retention of intricate relationships. (h) In contrast, the cPCA projection based on the classical kernel emphasizes dominant features, forming a more defined curve but potentially losing structural information. These distinct patterns highlight the differing strengths of qPCA and cPCA in preserving complex data structures during dimensional reduction. (i) An example of the quantum state vectors generated by the SAQK feature map, where all qubit states align near maximally coherent positions. This alignment reflects the successful utilization of quantum advantage in feature mapping, enabling superior representation of intricate data correlations.

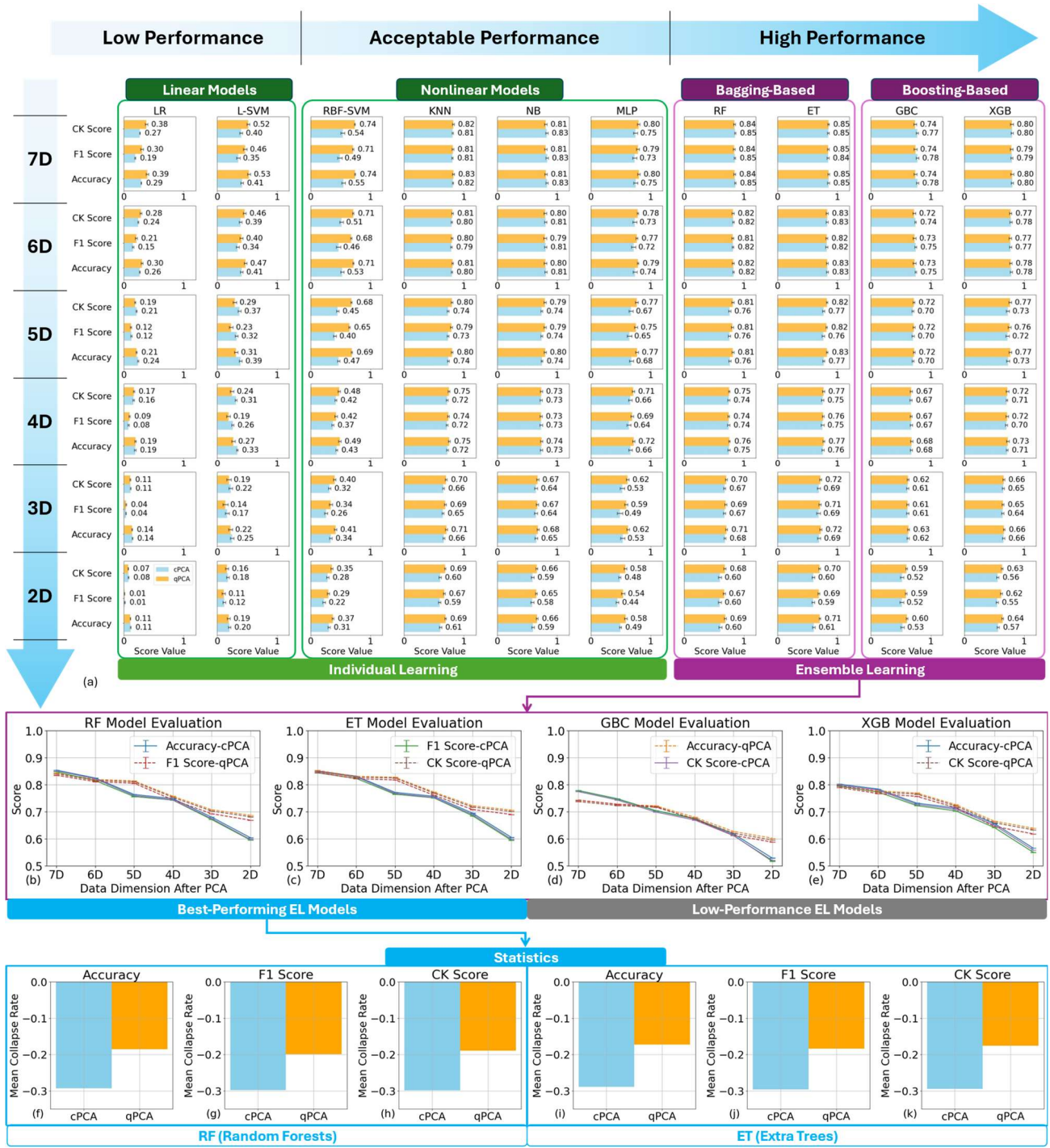

Fig. 3: Evaluation of ML model performance in different data spaces using classical cPCA and SAQK-based qPCA. The performance of representative ML models, including linear models (LR, L-SVM), non-linear models (RBF-SVM, KNN, NB, MLP), and ensemble learning (EL) models (RF, ET, GBC, XGB), is compared across various reduced dimensions (7D to 2D). (a) Key evaluation metrics (Accuracy, F1 Score, CK Score) are presented for all models under dimensional reduction. (b)-(e) Further evaluation focuses on the best-performing EL models, with the mean collapse rates (score reduction rates across data dimensions) extracted and analyzed in (f)-(k). These results demonstrate that the SAQK-based qPCA consistently achieves lower collapse rates across all three metrics compared to cPCA, highlighting its superior ability to retain critical information during dimensional reduction. These results suggest that qPCA preserves critical information more effectively during dimensional reduction, particularly for non-linear ML models, leading to superior evaluation scores across most dimensions. Detailed benchmarking of different quantum kernels is provided in Supporting Information.

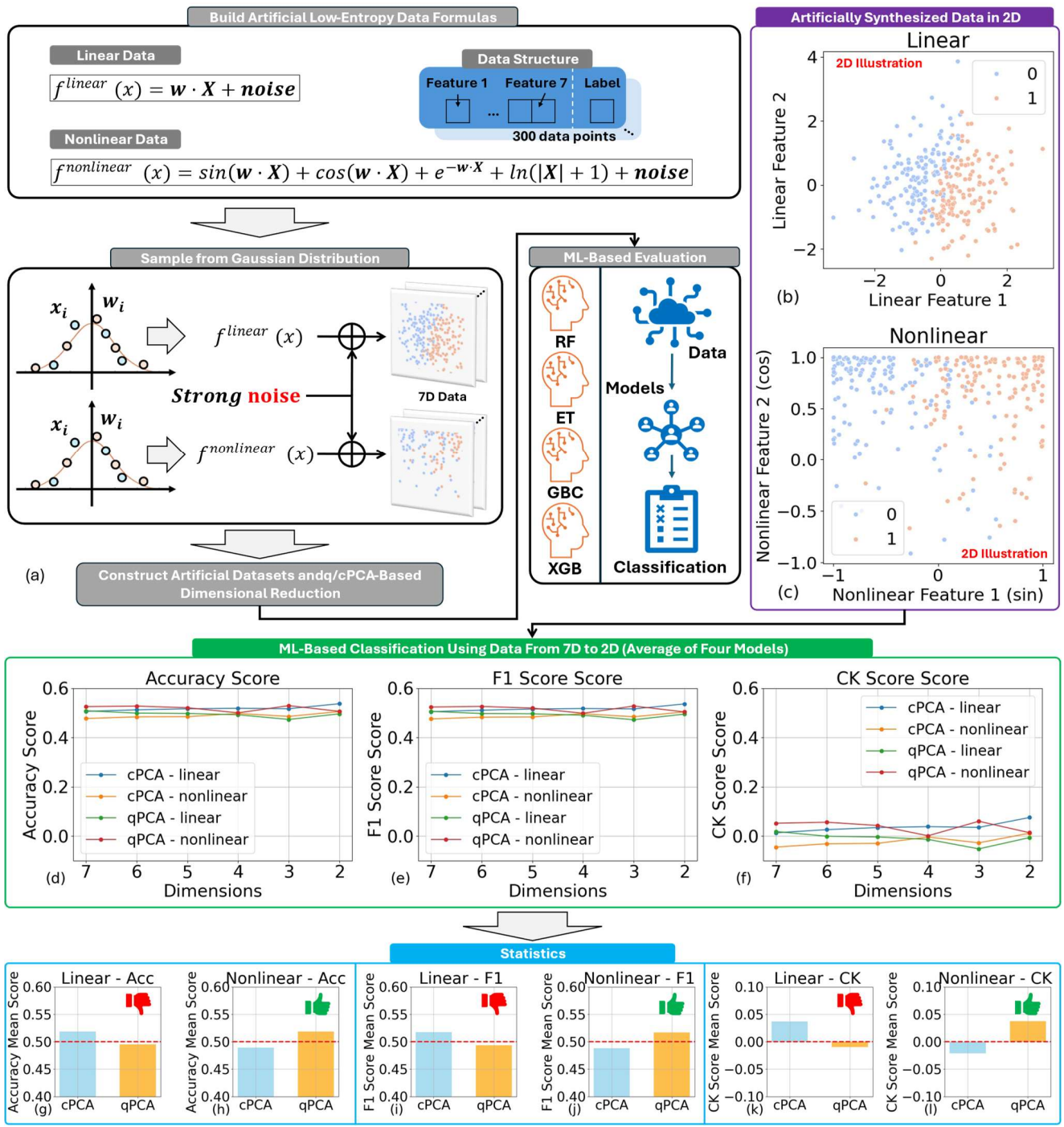

Fig. 4: Verification of SAQK in compressing artificially synthesized low-entropy datasets. (a) Workflow illustrating the synthesis of artificial linear and nonlinear low-entropy datasets for evaluating SAQK-based dimensional reduction. Each dataset is generated using parametric equations with inputs and weights both sampled from Gaussian distributions. To ensure sufficient complexity and realism, strong noise is introduced to mimic real-world scenarios and avoid bias toward EL models. The datasets are subsequently compressed and analyzed using four EL models: RF, ET, GBC, and XGB. (b)-(c) Visualization of synthesized data in 2D for clarity, with datasets for actual experiments containing 7 features to match the previous evaluation scenario. (d)-(f) Key evaluation scores (Accuracy, F1 Score, and CK Score) across reduced dimensions for different datasets, demonstrating consistent performance due to the low-entropy nature of the data. (g)-(l) Statistical analysis of evaluation scores (mean values) across dimensions highlights the advantage of SAQK in leveraging low-entropy data structures. Detailed experimental settings are provided in Methods.

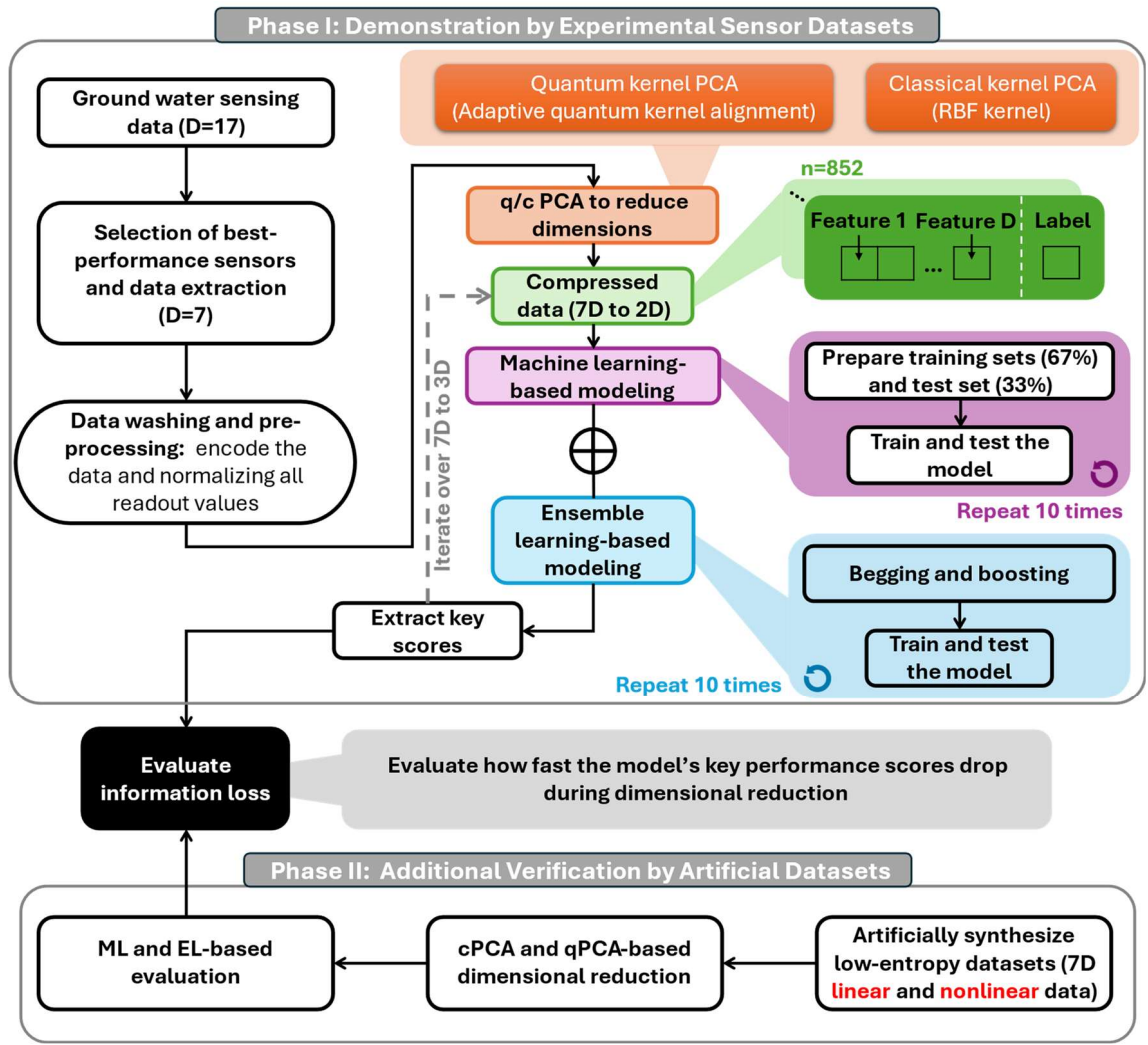

Fig. 5: Schematic flowline illustration of the procedure of this study. The Phase-I process aims to evaluate the performance of the proposed SAQK PCA in data compression. This phase begins with the collection of groundwater sensing data using 17 sensors (D=17). The best-performing sensors are selected by the method proposed in Ref. 28, reducing the data to 7 dimensions (D=7). The data is collected from a chemiresistive sensor array, which operates by detecting changes in resistance caused by chemical interactions between the sensing material and the target analytes in groundwater. Each sensor in the array is designed to respond to specific chemical properties. These interactions induce measurable variations in resistance ($\Delta R/R_0$), providing a multi-dimensional dataset that encapsulates both the chemical composition and environmental parameters of the water sample. The data is then washed and pre-processed by encoding and normalizing all readout values (see Supplementary Algorithm 2). Both quantum kernel PCA (fidelity kernel) and classical kernel PCA (RBF kernel) are applied to reduce the dimensions further (7D to 3D). Different quantum kernel methods are used for benchmarking. The compressed data is then split into training (80%) and test (20%) sets. Machine learning-based classification is then performed. The Phase-II process aims to verify the quantum advantage by applying the proposed SAQK PCA and the same evaluation on artificially synthesized low-entropy datasets. All evaluation results are averaged over the 10 repetitions to ensure reliability.